\documentclass[12pt]{article}
\usepackage{graphicx}
\usepackage{amssymb}
\usepackage{amsmath}

\newcommand{\bear}{\begin{array}}  
\newcommand {\eear}{\end{array}}
\newcommand{\bea}{\begin{eqnarray}}   
\newcommand{\eea}{\end{eqnarray}}
\newcommand{\beq}{\begin{equation}}   
\newcommand{\eeq}{\end{equation}}
\newcommand{\bef}{\begin{figure}}  \newcommand 
{\eef}{\end{figure}}
\newcommand{\bec}{\begin{center}}  \newcommand 
{\eec}{\end{center}}

\def\lrfp#1#2#3{ \left(\frac{#1}{#2} 
\right)^{#3}}

\def\lrfp#1#2#3{ \left(\frac{#1}{#2} 
\right)^{#3}}

\begin{document}

\begin{titlepage}

\begin{flushright}
IPMU 08-0072 \\
ICRR-Report-530
\end{flushright}

\vskip 1.35cm

\begin{center}
{\large \bf
Hilltop Non-Gaussianity
}
\vskip 1cm

Masahiro Kawasaki$^{(a,b)}$,
Kazunori Nakayama$^{(a)}$ and
Fuminobu Takahashi$^{(b)}$

\vskip 0.4cm

{ \it $^a$Institute for Cosmic Ray Research,
University of Tokyo, Kashiwa 277-8582, Japan}\\
{\it $^b$Institute for the Physics and Mathematics of the Universe,
University of Tokyo, Kashiwa 277-8568, Japan}
\date{\today}

\begin{abstract}
We study non-Gaussianity induced by a pseudo Nambu-Goldstone boson
with a cosine-type scalar potential. We focus on how the
non-Gaussianity is affected when the pseudo Nambu-Goldstone boson
rolls down from near the top of the scalar potential where
the deviation from a quadratic potential is large. We find that
the  resultant non-Gaussianity is similar to that obtained in the quadratic potential,
if the pseudo Nambu-Goldstone boson accounts for the curvature perturbation;
the non-Gaussianity is enhanced, otherwise.
\end{abstract}

%\preprint{IPMU-08-???}
%\preprint{ICRR-Report-???}
%\pacs{98.80.Cq}

\end{center}
\end{titlepage}

%%%%%%%%%%%%%%%%%%%%%%%%%%%%%%%%%%%%
\section{Introduction}
%%%%%%%%%%%%%%%%%%%%%%%%%%%%%%%%%%%%
Currently, non-Gaussianity of the cosmological perturbation is  a hot topic,
since Yadav and Wandelt reported a detection of large non-Gaussianity in the WMAP 3yr data on
the cosmic microwave background (CMB) temperature anisotropy~\cite{Yadav:2007yy}.
The latest WMAP 5yr data is consistent with vanishing non-Gaussianity at 95\% C.L.~\cite{Komatsu:2008hk},
although the data still seems to favor similar deviation from the pure Gaussian statistics.
It is not yet settled whether there are non-Gaussianities, in spite of many recent activities
searching for their observational hints \cite{Slosar:2008hx}.

On the theoretical side, to generate an observable amount of non-Gaussianity requires
significant modifications on the standard single-field slow-roll inflation scenario,
where all the density perturbations come from the quantum fluctuation of the inflaton.
Perhaps the simplest modification is to introduce an additional light scalar field, 
since the density perturbation of such a scalar field always contains some amount of non-Gaussianity.
For instance, a curvaton field, which accounts for the curvature perturbations, can generate
large non-Gaussianity \cite{Mollerach:1989hu,Linde:1996gt,Lyth:2001nq}.
It is also possible to generate large non-Gaussianity with negligible contribution to the curvature perturbation, 
and we call such a  scalar as an ungaussiton \cite{Suyama:2008nt}.
Moreover, a light scalar may be stable and contributes to some fraction of the dark matter of the Universe,
as in the case of axion \cite{Peccei:1977hh}.
In this case, the axion has a CDM isocurvature perturbation, and as a consequence,
the non-Gaussianity of isocurvature-type is generated
\cite{Bartolo:2001cw,Boubekeur:2005fj,Kawasaki:2008sn,Langlois:2008vk}.

Non-Gaussianity from a light scalar, particularly in the curvaton scenario,
has been extensively studied in the literature \cite{Lyth:2002my,Lyth:2005fi},
but most of them considered a simple quadratic potential or
a potential slightly modified from a quadratic one \cite{Enqvist:2005pg,Huang:2008bg}.
In this paper we study how the resultant non-Gaussianity is affected in a case that the scalar potential is
largely deviated from the quadratic potential.
One example is a pseudo Nambu-Goldstone (pNG) boson, and the other is
a scalar rolling from the origin in a symmetry breaking potential.
The former example is particularly interesting, since
the flatness of the potential is protected by the symmetry,
which enables the pNG boson to have large quantum fluctuation.
Thus a pNG boson is a good candidate for the curvaton \cite{Dimopoulos:2003az}.
Thus it is worth studying the non-Gaussianity arising from the fluctuation of the pNG boson,
which has a cosine-type potential.
Despite its importance, less attention was paid to the non-Gaussianity in the pNG curvaton model.
In particular, if the pNG boson stays initially around the top of the potential during inflation, 
the predictions on non-Gaussianity are expected to change 
compared with those in the standard curvaton scenario with
a simple quadratic potential and such a case was not studied in the literature, so far.

This paper is organized as follows.
In Sec.~\ref{sec:NL} a general formalism for calculating the non-Gaussainity in the curvaton scenario
is briefly reviewed.
In Sec.~\ref{sec:NGSB} we study the non-Gaussianity from pNG type curvaton scenario.
Sec.~\ref{sec:conc} is devoted to conclusions.

%%%%%%%%%%%%%%%%%%%%%%%%%%%%%%%%%%%%
\section{Non-linearity Parameter} \label{sec:NL}
%%%%%%%%%%%%%%%%%%%%%%%%%%%%%%%%%%%%

According to the $\delta N$ formalism \cite{Starobinsky:1986fxa,Sasaki:1995aw,Lyth:2004gb},
the curvature perturbation $\zeta$ on sufficiently large scales 
on the uniform density slicing is evaluated by 
\begin{equation}
	\zeta = N(t_i,t_f; \vec x) - \bar N(t_i,t_f),
\end{equation}
where $N(t_i,t_f; \vec x)$ is the $e$-folding number from the initial spatially flat hypersurface at $t=t_i$,
to the final uniform density slicing at $t=t_f$ along the world line $\vec x =$const., and 
$\bar N(t_i,t_f)$ is the background $e$-folding number. We take 
the initial flat slicing at a time when a scale of interest exits the horizon during inflation,
and set the final uniform density slicing in the final radiation dominated epoch before the 
cosmological scales re-enter the horizon. 
$\zeta$ is known to be a gauge-invariant quantity and time-independent in the absence of 
non-adibatic pressure \cite{Wands:2000dp}.
Assuming that there are light scalar fields, $\{ \phi_a \}$, during inflation,
$\zeta$ can be expanded as
\begin{equation}
	\zeta = N_a \delta \phi_a + \frac{1}{2}N_{ab} \delta \phi_a \delta \phi_b + \dots ,
\end{equation}
where $\delta \phi_a$ denotes the fluctuation of $\phi_a$ during inflation at $t=t_i$,
and $N_a = \partial N/\partial \phi_a$, and so on.
The power spectrum and the bispectrum of $\zeta$ are written as
\bea
	\langle \zeta_{\vec k_1}\zeta_{\vec k_2} \rangle 
	&=&(2\pi)^3\delta(\vec k_1+\vec k_2) P_\zeta(k_1), \\
	\langle \zeta_{ {\vec k_1}}\zeta_{ {\vec k_2}}\zeta_{ {\vec k_3}} \rangle
	&=&(2\pi)^3\delta(\vec k_1+\vec k_2+\vec k_3)B_\zeta (k_1,k_2,k_3).
\eea
The non-linearity parameter $f_{\rm NL}$ is defined by 
\begin{equation}
	B_\zeta (k_1,k_2,k_3) = \frac{6}{5}f_{\rm NL} \left[ P_\zeta(k_1) P_\zeta(k_2)
	+P_\zeta(k_2) P_\zeta(k_3) + P_\zeta(k_3) P_\zeta(k_1) \right].
\end{equation}
In a squeezed configuration such that one of the $k_i$'s $(i=1,2,3)$ is much smaller
than the other two, $f_{\rm NL}$ can be expressed in terms of $N_a$ and $N_{ab}$ as \cite{Lyth:2005fi}
\begin{equation}
	\frac{6}{5}f_{\rm NL} = \frac{N_aN_bN_{ab}+N_{ab}N_{bc}N_{ca}\Delta_{\delta \phi}^2
	\ln(k_b L) }{(N_aN_a)^2},
\end{equation}
where  $k_b \equiv{\rm min}\{k_1,k_2,k_3\}$, and we have
 introduced an infrared cutoff $L$, which is taken to be a few times larger 
than the present Hubble horizon scale \cite{Lyth:1991ub,Lyth:2007jh}.
We have here defined
\beq
	\Delta_{\delta \phi}^2 \;\equiv\; \frac{k^3}{2\pi^2} P_{\delta \phi}(k)
	\simeq \lrfp{H_{\rm inf}}{2\pi}{2},
\eeq
where $H_{\rm inf}$ is the Hubble parameter during inflation.

Let us consider a simple curvaton model where a light scalar field $\sigma$ is responsible for
the total curvature perturbation, and the inflaton $\phi$ has negligible contributions to the bispectrum.
Then the expression is simplified to
\begin{equation}
	\frac{6}{5}f_{\rm NL} = \frac{N_\sigma^2 N_{\sigma \sigma}
	+N_{\sigma \sigma}^3 \Delta_{\delta \phi}^2 \ln(k_b L) }
	{(N_\phi^2+N_\sigma^2)^2}.  \label{fNL}
\end{equation}
Assuming that the oscillating $\sigma$ field behaves like non-relativistic matter,
$N_\sigma$ and $N_{\sigma \sigma}$ are given by \cite{Lyth:2005fi}
\begin{equation}
	N_\sigma = \frac{2R}{3}\frac{\sigma_{\rm os}'}{\sigma_{\rm os}}, ~~~
	N_{\sigma \sigma}=\frac{2R}{3}
	\left[\left(1+\frac{\sigma_{\rm os}\sigma_{\rm os}''}{\sigma_{\rm os}^{\prime 2}} \right)
	-\frac{4}{3}R-\frac{2}{3}R^2\right ]\left(\frac{\sigma_{\rm os}'}{\sigma_{\rm os}} \right)^2,
\end{equation}
where $\sigma_{\rm os}$ denotes the field value at the onset of its oscillation 
and the prime denotes a derivative with respective to the initial position at $t=t_i$,
denoted by $\sigma_i$, and $R$ roughly denotes the fraction of the curvaton energy density at the time of its decay,
\begin{equation}
	R = \left (\frac{3\rho_\sigma}{4\rho_r +3\rho_\sigma} \right )_{\rm \sigma~decay}.
\end{equation}
In the case of quadratic potential, $\sigma_{\rm os} =\beta \sigma_i$ 
with an $\mathcal O(1)$ constant %$\beta=2^{1/4}\Gamma(5/4)J_{1/4}(1)$ 
$\beta=(2/mt_{\rm os})^{1/4}\Gamma(5/4)\\ \times J_{1/4}(mt_{\rm os})$ 
($J_n(x)$ is the Bessel function) where $t_{\rm os}$ denotes the epoch at which $\sigma$
begins to oscillate,
and hence $\sigma_{\rm os}'/\sigma_{\rm os}=1/\sigma_i$.
If the curvaton potential receives correction from non-quadratic terms, 
the resulting non-Gaussianity may significantly change \cite{Enqvist:2005pg}.
Assuming that the curvaton potential is quadratic
and the curvaton fluctuation is dominated by the linear part, 
i.e., $|\delta \sigma / \sigma_i | \ll 1$, we obtain
\begin{equation}
	f_{\rm NL} = \frac{5}{4R}\left (1- \frac{4}{3}R-\frac{2}{3}R^2 \right ), \label{fNL-R}
\end{equation}
where we have neglected the contribution to the curvature perturbation from the inflaton fluctuation.
Thus a small value of $R$ is necessary to generate large non-Gaussianity.
Note that the above discussion relied on an assumption that the curvaton behaves like
non-relativistic matter, and this means that the curvaton potential is (almost) quadratic.
However, we are interested in a case where the curvaton oscillates in the cosine-type
potential, and the above formula cannot be applied to such a case.
In the next section we study the non-Gaussianity in the pNG curvaton scenario.

%%%%%%%%%%%%%%%%%%%%%%%%%%%%%%%%%%%%%%%%%
\section{Non-Gaussianity from pseudo Nambu-Goldstone boson}  \label{sec:NGSB}
%%%%%%%%%%%%%%%%%%%%%%%%%%%%%%%%%%%%%%%%%

As is well known, a large non-Gaussianity can be obtained in the curvaton scenario, only if $R \ll 1$.
However, as already noted in the introduction, one of the difficulties to realize the curvaton scenario is that 
scalar fields other than the inflaton generically have masses of order of the Hubble scale in supergravity
and hence their quantum fluctuations are highly suppressed.
In order to forbid such Hubble mass terms, one needs some special inflation model
such as the $D$-term inflation model \cite{Binetruy:1996xj},
or tuning of the coupling between the curvaton and inflaton.
One of the most promising ways to achieve this is to identify the curvaton as 
a pNG boson associated with spontaneous symmetry breaking,
since it does not receive a Hubble mass correction.
Therefore a pNG boson is a good candidate for the curvaton \cite{Dimopoulos:2003az}.

Consider a case that a complex scalar $\Phi$ has a vacuum expectation value and spontaneously breaks 
U(1) symmetry.
If the symmetry is exact, a NG boson corresponding to the angular direction of $\Phi$, 
denoted by $\sigma$, remains massless.
But if the symmetry is explicitly broken by a small amount, $\sigma$ feels 
an axion-type potential with a symmetry breaking scale $f$,
\begin{equation}
	V(\sigma) \;=\;m^2 f^2 \left[ 1 - \cos \left ( \frac{\sigma}{f} \right ) \right].
\end{equation}
where $m$ is the mass of $\sigma$ field.

\begin{figure}[t]
 \begin{center}
 \includegraphics[width=0.6\linewidth]{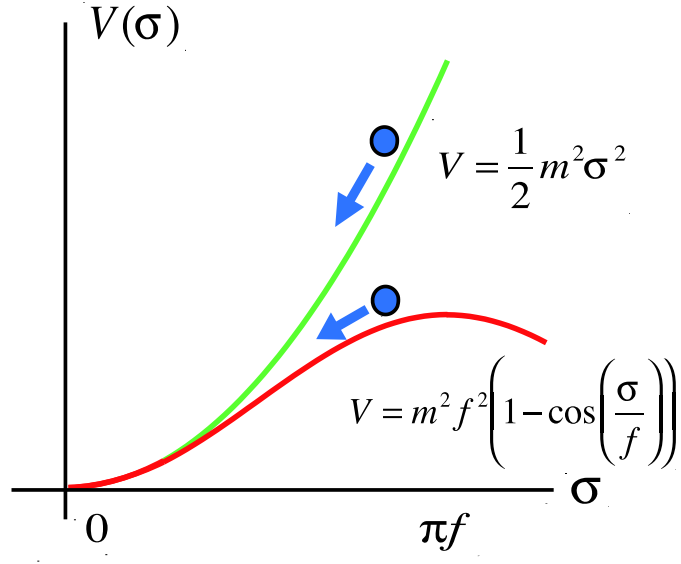}
  \caption{
  Schematic pictures of the quadratic and axion-type potential.
  }
   \label{fig:pot}
 \end{center}
\end{figure}

Let us suppose that the symmetry is already broken during inflation, 
and is not restored after that.
Then $\sigma$ freezes at the initial value during inflation ($\sigma_i$) until it starts to oscillate.
In principle $\sigma_i$ can take any value between $0$ and $2\pi f$.
If $\sigma$ happens to sit near the origin ($|\sigma_i | \ll \pi f/2$), its potential is approximated by
\begin{equation}
	V(\sigma) \;\simeq\; \frac{1}{2}m^2 \sigma^2,  \label{quad}
\end{equation}
and hence it reproduces the usual results of the curvaton models with a quadratic potential
(see Fig.~\ref{fig:pot}).
However, $\sigma$ may be relatively close to the top of the potential where the above approximation 
breaks down.
This case was not studied in previous works.
One complexity arises due to a fact that the oscillating $\sigma$ does not behave like 
non-relativistic matter since the potential is significantly deviated from the quadratic one in this case.

%%%%%%%%%%%%%%%%%%FIGURE%%%%%%%%%%%%%%%%%%%

%%%%%%%%%%%%%%%%%%%%%%%%%%%%%%%%%%%%%%%%%

We have performed a numerical calculation to evaluate $N_\sigma$ and $N_{\sigma \sigma}$.
We have solved a set of equations
\begin{gather}
	\ddot{\sigma} + 3H\dot \sigma = -\frac{\partial V}{\partial \sigma}, \\
	H^2 = \frac{1}{3M_P^2}\left ( \rho_r + \rho_\sigma \right ), \\
	\rho_\sigma = \frac{1}{2}{\dot \sigma}^2 + V(\sigma).
\end{gather}
from the initial time $t_i$, taken to be well before $\sigma$ begins to oscillate,
to the final time $t_f$, taken to be the time of $\sigma$ decay, under a sudden decay approximation.
For various initial conditions $\sigma_i$, we  calculate the total $e$-folding number 
$N(\sigma_i)$, and its first and second derivative, $N_\sigma$ and $N_{\sigma \sigma}$.
%Then using Eq.~(\ref{fNL}) we  calculate $f_{\rm NL}$
%(here we neglect the second term of the r.h.s. of Eq.~(\ref{fNL})).
We calculate $f_{\rm NL}$ using Eq.~(\ref{fNL}), where we assume that $\sigma_i$ is chosen
in such a way that the first term dominates over the second one in Eq.~(\ref{fNL}).
As $\sigma_i$ approaches the top of the potential for fixed $H_{\rm inf}$, the $e$-folding number $N$ 
diverges since $\sigma$ stays there for longer and longer duration in that limit
and the perturbative expansion breaks down.\footnote{In reality, the $e$-folding number
does not go to the infinity, and instead, domain walls will appear, making the estimate
on the non-Gaussianity more complicated.}

One may expect that the sign of $f_{\rm NL}$ changes around $\sigma_i =\pi f/2$,
reflecting the change of the sign of the second derivative of the potential.
The naive estimation goes as follows.
Noting that the curvature perturbation generated by the curvaton $\sigma$ is given by
\begin{equation}
	\zeta_\sigma \simeq \frac{1}{3}\frac{\delta \rho_\sigma}{\rho_{\rm tot}}
	\simeq \frac{R}{3}\left[ \frac{V'}{V}\delta \sigma +\frac{V''}{2V}(\delta \sigma)^2 \right],	
\end{equation}
where $\rho_{\rm tot}$ is the total energy density of the Universe at
the curvaton decay and the prime denotes the derivative with respective to $\sigma$, we naively expect 
\begin{equation}
	N_\sigma \simeq \frac{R}{3}\frac{\sin (\sigma_i/f)}{f[1-\cos (\sigma_i/f)]},~~~
	N_{\sigma \sigma} \simeq \frac{R}{3}\frac{\cos (\sigma_i/f)}{f^2[1-\cos (\sigma_i/f)]}.
\end{equation}
Thus using Eq.~(\ref{fNL}), we would obtain
\begin{equation}
	f_{\rm NL}\simeq \frac{5}{2R}\frac{\cos (\sigma_i/f)[1-\cos (\sigma_i/f)]}
	{\sin ^2(\sigma_i/f)}.
\end{equation}
In the limit $\sigma_i \ll f$, the result for the quadratic potential $f_{\rm NL}\simeq 5/(4R)$
is reproduced.
On the other hand, this expression shows that $f_{\rm NL}$ changes the sign
for $\sigma_i > (\pi/2)f$.
%Near the top of the potential, we can approximate $V \sim m^2 f^2 - m^2 (\Delta \sigma)^2/2 $
%where $\Delta \sigma \equiv \pi f- \sigma$. 
%Thus the fractional fluctuation of the curvaton energy density reads
%%
%\begin{equation}
%	\frac{\delta \rho_\sigma}{\rho_\sigma}\simeq 
%	\frac{-\Delta \sigma \delta \sigma - (\delta \sigma)^2 }{f^2},
%\end{equation}
%%
%hence we may expect the negative non-linearity parameter given by
In particular, if the curvaton is initially placed near the top of the potential, we would obtain
$f_{\rm NL} \simeq - (f/\Delta \sigma)^2(5/R)$,
where $\Delta \sigma \equiv \pi f- \sigma_i$.
However, note that this expression ignores the kinetic energy, and
we show that this naive estimation is not correct.
%\footnote{
%	If $|\Delta \sigma| < |\delta \sigma|$, domain walls are formed due to the fluctuation of the 
%	$\sigma$ field, which may lead to cosmological disaster.
%	We do not consider such a case.
%}

%%%%%%%%%%%%%%%%%%FIGURE%%%%%%%%%%%%%%%%%%%

\begin{figure}[]
 \begin{center}
 \includegraphics[width=0.5\linewidth]{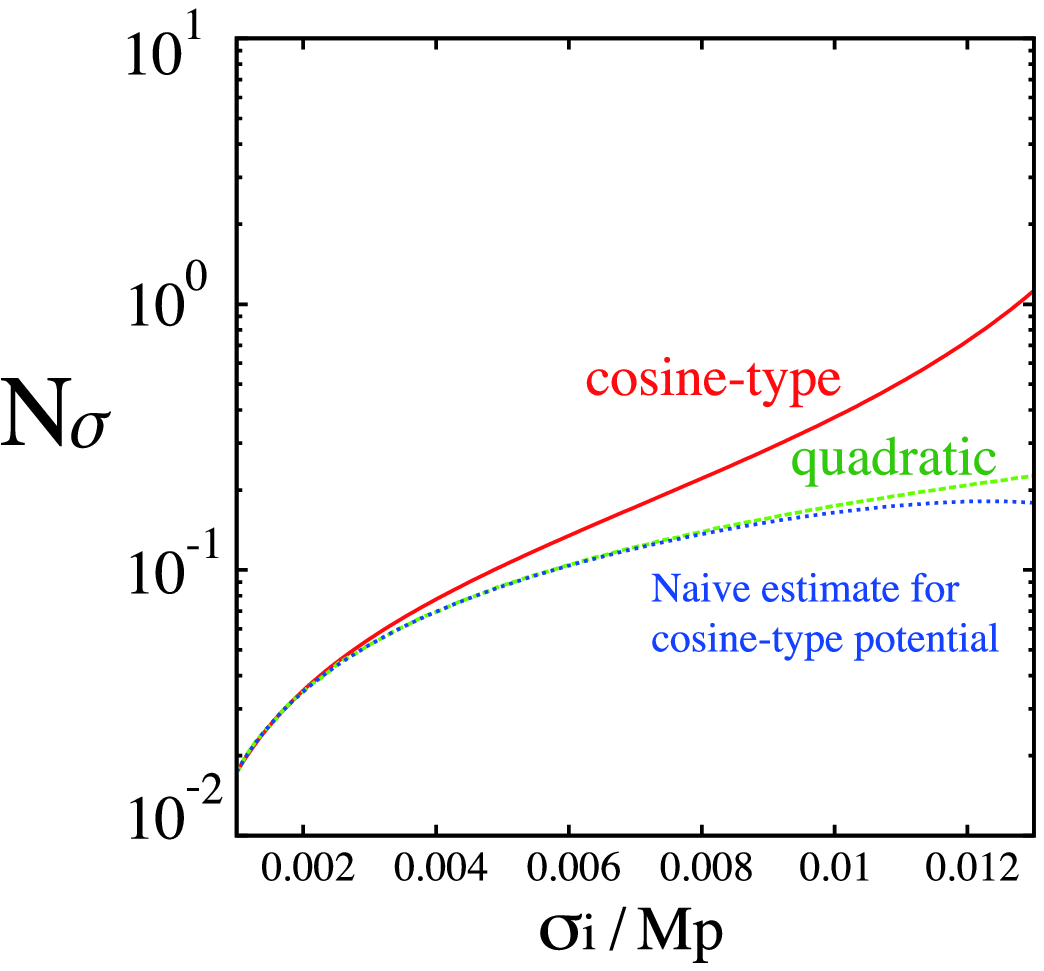}
 \includegraphics[width=0.5\linewidth]{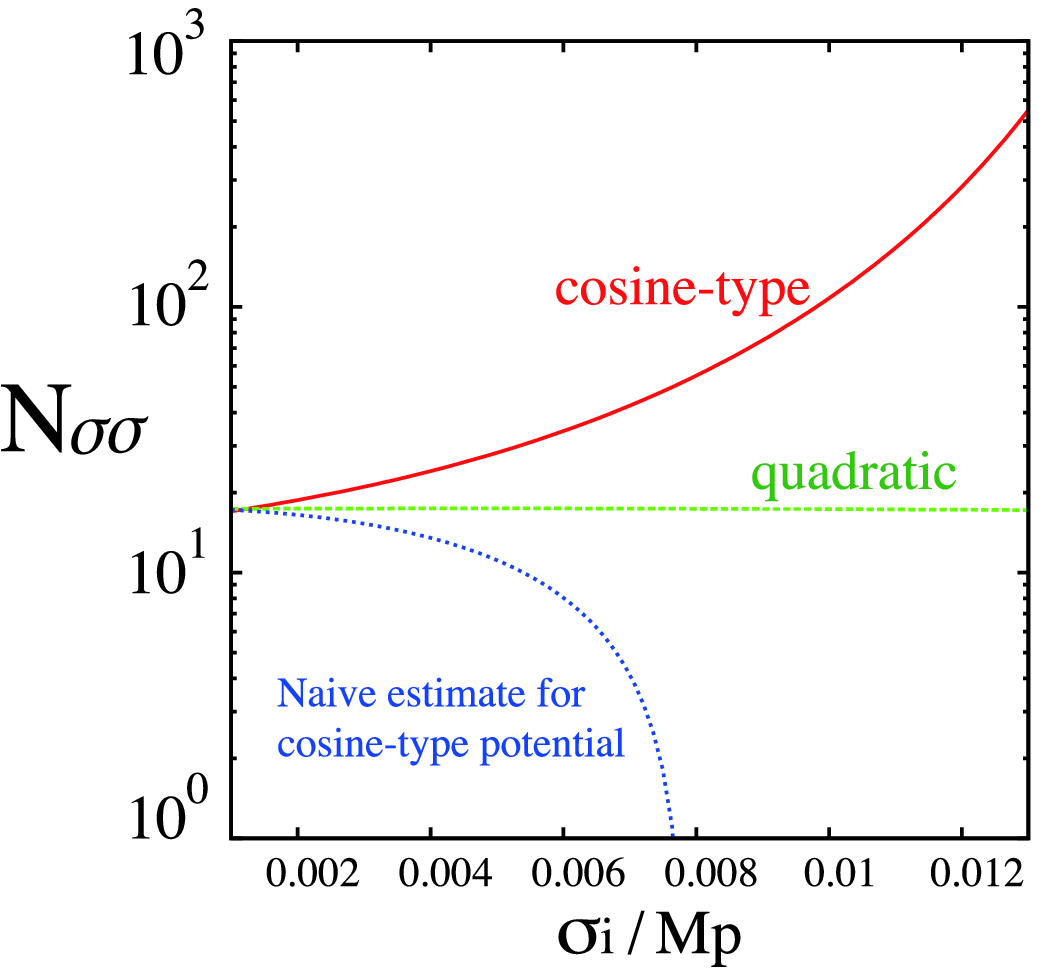}
   \caption{
  $N_\sigma$ (top) and $N_{\sigma \sigma}$ (bottom) in Planck units
   as a function of initial position of the curvaton, $\sigma_i$,
   for the quadratic and cosine-type potentials.
   Here we have set $f=5\times 10^{-3}M_P$.
   We also show the naive estimation of $f_{\rm NL}$ for the cosine-type potential.
  }
   \label{fig:Nsig}
 \end{center}
\end{figure}

%%%%%%%%%%%%%%%%%%%%%%%%%%%%%%%%%%%%%%%%%

\begin{figure}[t]
 \begin{center}
 \includegraphics[width=1.0\linewidth]{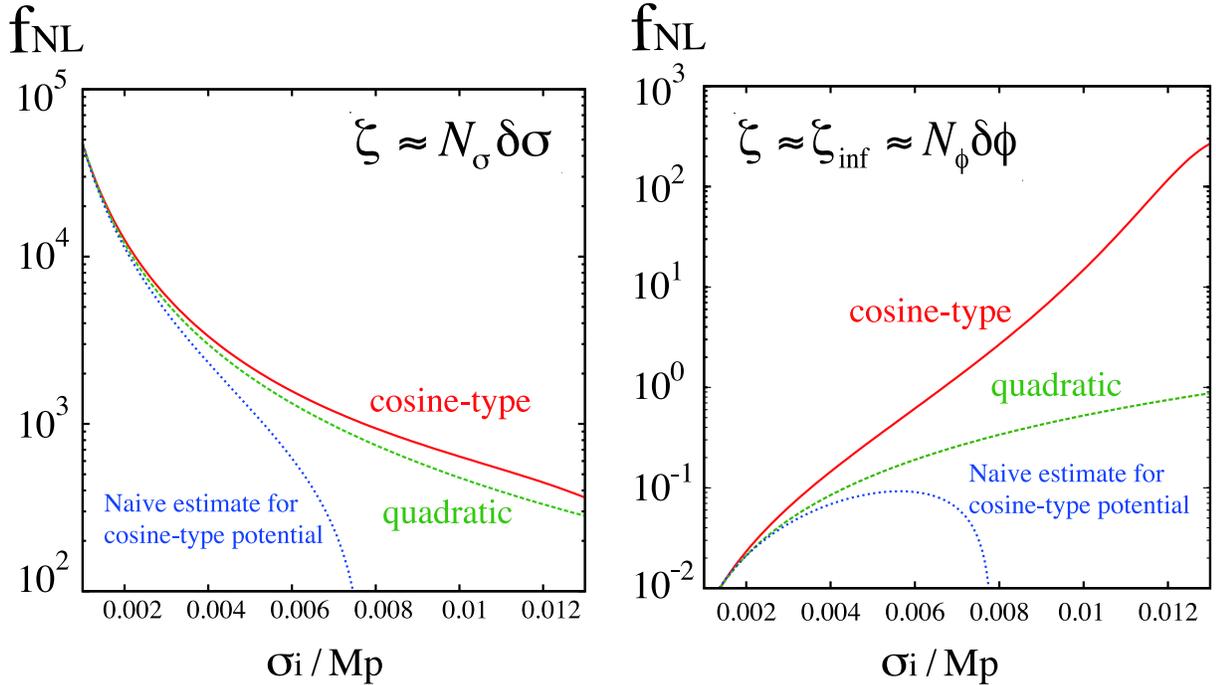}
   \caption{
   The value of non-linerity parameter $f_{\rm NL}$ 
   as a function of initial position of the curvaton, $\sigma_i$,
   for the quadratic and cosine-type potentials,
   for $N_\sigma \gg N_\phi$ (left) and $N_\phi=1$ (right).
   Here we have set $f=5\times 10^{-3}M_P$ and $\Gamma_\sigma = 10^{-4}m$.
   We also show the naive estimation of $f_{\rm NL}$ for the cosine-type potential.
   The value of $f_{\rm NL}$ decreases (increases) 
   as $\Gamma_\sigma$ becomes smaller for the left (right) panel,
   although the relation between the three lines will not change.
  }
   \label{fig:fnl}
 \end{center}
\end{figure}

The results of the numerical calculation are shown in Figs.~\ref{fig:Nsig} and \ref{fig:fnl}.
In Fig.~\ref{fig:Nsig} we have plotted $N_\sigma$ and $N_{\sigma \sigma}$ in Planck units,
as a function of the initial position $\sigma_i$.
The left panel of Fig.~\ref{fig:fnl} shows $f_{\rm NL}$ calculated from Eq.~(\ref{fNL}) assuming 
that the curvature perturbation is dominantly generated by the curvaton ($N_\sigma \gg N_\phi$).
For comparison, we have also shown the numerical result for the quadratic potential (\ref{quad}).
Here we have set the symmetry breaking scale as $f=5\times 10^{-3}M_P$ and
the decay rate of $\sigma$ as $\Gamma_\sigma=10^{-4}m$,
and assumed that the reheating temperature after inflation is so high that the
Universe is dominated by the radiation when $\sigma$ starts to oscillate.
It is seen that the value of $N_\sigma$ and $N_{\sigma \sigma}$ for the cosine-type potential
significantly deviate from the quadratic one,
and also from the naive estimation for the cosine-type potential.\footnote{
	In order to reproduce the observed density perturbation, the inflationary scale
	should take an appropriate value for each point.
	For example, $N_\sigma \lesssim 1$ requires 
	unacceptably large Hubble scale during inflation, $H_{\rm inf}\gtrsim 10^{14}$~GeV.
	However, this is just an artifact of our choice of $\Gamma_\sigma$ for our numerical calculation.
	The inflationary scale can be sufficiently small if we take a smaller value of $\Gamma_\sigma$.
}
Surprisingly, however, the resulting non-Gaussianity for the cosine-type potential is very similar to
that of the quadratic one, irrespective of the initial field value.
This is due to an accidental cancellation between $N_\sigma^2$ and $N_{\sigma \sigma}$
in (\ref{fNL}).
Actually, the sign of $f_{\rm NL}$ is always {\it positive} even if the initial position of the curvaton field is
located near the top of the potential.
This does not match with the previous naive estimation.
The resolution lies on an implicit assumption that the curvaton begins to oscillate at $H=m$ and afterwards
it behaves exactly like non-relativistic matter.
Both are incorrect in general.
If the initial position of the curvaton is near the top of the potential, the oscillation epoch is
significantly delayed, which causes more contribution to the $e$-folding number,
leading to the enhancement in $N_\sigma$ and 
$N_{\sigma \sigma}$ (see Fig.~\ref{fig:Nsig}).
Therefore the positive $f_{\rm NL}$ is obtained even in this case.
This should be contrasted to the case studied in Ref.~\cite{Enqvist:2005pg}.
They studied the curvaton potential with a small positive correction to the quadratic potential,
and obtained a negative value of $f_{\rm NL}$.
This can be understood as a result of the suppression of the growth of the $e$-folding number
due to the positive correction term.\footnote{
	It was noticed in Ref.~\cite{Huang:2008bg} that a negative correction to the 
	quadratic potential leads to an enhancement on $f_{\rm NL}$,
	in the case that the correction term is small enough and can be treated as a perturbation.
}

%%%%%%%%%%%%%%%%%%FIGURE%%%%%%%%%%%%%%%%%%%

%%%%%%%%%%%%%%%%%%%%%%%%%%%%%%%%%%%%%%%%%

In the above arguments, we have assumed that the curvature perturbation is dominantly
generated by the pNG boson. The situation changes if
the inflaton accounts for the curvature perturbation.
In the right panel of Fig.~\ref{fig:fnl} we show the non-linearity parameter $f_{\rm NL}$ 
in the presence of the contribution of the inflaton to the curvature perturbation,
setting $N_\phi=1$ in Planck units.
In this case, the non-Gaussianity in the cosine-type potential can be orders of magnitude larger than 
that in the quadratic potential.

This is a good news for those who want to produce large non-Gaussianity in the curvaton scenario,
since one needs not to tune the initial value of the curvaton field.
Once there are pNG fields during inflation, it always predicts positive value of $f_{\rm NL}$
irrespective of its initial position.
Of course, in order to obtain observable level of non-Gaussianity, $R$ should be smaller than 
about 0.1, and it requires a tuning of the decay rate of $\sigma$.
However, in the pNG scenario both the amplitude and the decay rate of the curvaton
can be in principle determined by an underlying physics, not by rather arbitrary initial conditions.
Actually the decay rate of the curvaton is expected to be $\Gamma_\sigma \sim m^3/f^2$
and its initial amplitude is $\sim f$.
Thus $R$ can be estimated as
\begin{equation}
	R \sim 0.3\left ( \frac{T_R}{10^5~{\rm GeV}} \right )
	\left ( \frac{f}{10^{15}~{\rm GeV}} \right )^3
	\left ( \frac{1~{\rm TeV}}{m} \right )^{3/2},
\end{equation}
where $T_R$ is the reheating temperature after inflation.
Although it may be difficult to distinguish such a scenario from standard curvaton models,
it is important to keep in mind that the detection of non-Gaussianity may be closely related to
an underlying high-energy physics.

%%%%%%%%%%%%%%%%%%%%%%%%%%%%%%%%%%%%
\section{Discussion and Conclusions}  \label{sec:conc}
%%%%%%%%%%%%%%%%%%%%%%%%%%%%%%%%%%%%

In this letter we have investigated the non-Gaussianity from a light scalar with a potential
largely deviated from the quadratic one, focusing on
the pNG curvaton scenario,
motivated by the fact that the required flatness of the potential is naturally ensured.
What we have found is that if the potential significantly deviates from the quadratic one,
naive estimation on the non-linearity parameter $f_{\rm NL}$ is not correct in general.
This is because the timing when the oscillation begins and the equation of state during the 
coherent oscillations are different from those in the quadratic potential. Interestingly,
if the curvaton accounts for the observed curvature perturbations,
 the non-linearity parameter is quite similar to that of the usual quadratic one,
irrespective of the initial position of the curvaton field. If the inflaton is responsible for the
curvature perturbation, the non-Gaussianity is enhanced in the pNG curvaton case.
Therefore, the pNG curvaton scenario is a good example for realizing a large non-Gaussianity.

Although we have focused on the pNG potential, a similar analysis can be applied to other
hilltop potentials.
For example, consider the potential given by
\begin{equation}
	V(\sigma)=\frac{1}{2}m^2 \sigma^2 - \frac{\lambda}{4} \sigma^4 ,
\end{equation}
with self-coupling constant $\lambda (>0)$, which has an extremum at $\sigma = m/\sqrt{\lambda}$.
If the $\sigma$ field rolls down toward the origin from near this extremum, 
the resultant non-Gaussianity is similar to that predicted in the corresponding quadratic potential,
as in the pNG case.
Therefore our results indicate that if the curvaton has  a similar hilltop potential which changes to the quadratic
one at the potential minimum, the non-Gaussianity can be estimated approximately by the corresponding quadratic potential,
if $\sigma$ is dominantly responsible for the total curvature perturbation.

Finally we comment on the non-Gaussianity generated by the axion.
In Ref.~\cite{Kawasaki:2008sn}, non-Gaussianity from the axion was studied
focusing on the case that the axion initially stays near the potential minimum 
and the potential is well approximated by the quadratic one.
According to the result of this paper, if the axion is placed near the top of the potential,
the resultant non-Gaussianity as well as its isocurvature perturbation are enhanced.

%%%%%%%%%%%%%%%%%%%%%%%%%%%%%%%%%%%%
\section*{Acknowledgment}
%%%%%%%%%%%%%%%%%%%%%%%%%%%%%%%%%%%%

%%%%%%%%%%%%%%%%%%%%%%%%%%%%%%%%%%%%%%%%%%%%
%\begin{acknowledgements}
%%%%%%%%%%%%%%%%%%%%%%%%%%%%%%%%%%%%%%%%%%%%

K.N. would like to thank the Japan Society for the Promotion of Science for financial support.
This work is supported by Grant-in-Aid for Scientific research from the Ministry of Education,
Science, Sports, and Culture (MEXT), Japan, No.14102004 (M.K.)
and also by World Premier International
Research Center InitiativeiWPI Initiative), MEXT, Japan.

 %%%%%%%%%%%%%%%%%%%%%%%%%%%%%%%%%%%%%%%%%%%%
%\end{acknowledgements}
%%%%%%%%%%%%%%%%%%%%%%%%%%%%%%%%%%%%%%%%%%%%

%%%%%%%%%%%%%%%%%%%%%%%%%%%%%%%%%%%%%%%%%%%%

%%%%%%%%%%%%%%%%%%%%%%%%%%%%%%%%%%%%%%%%%%%%

\end{document}